\documentstyle[epsfig,spie]{article} 

\title{Imaging the Universe in 3D with the VLT: 
The Next Generation Field Spectrometer SPIFFI} 

\author{Frank Eisenhauer, Matthias Tecza, Sabine Mengel, Niranjan
Thatte, \\
Claudia R\"ohrle, Klaus Bickert and J\"urgen Schreiber
\skiplinehalf 
Max-Planck-Institut f\"ur extraterrestrische Physik,
Giessenbachstrasse, 85740 Garching, Germany 
}

\authorinfo{E-mail: eisenhau@mpe.mpg.de}

\begin{document} 
\maketitle 

\begin{abstract}

We present SPIFFI, the integral field spectrometer for the VLT. 
This instrument allows simultaneous observation of infrared spectra in 
more than 1000 image points of a two dimensional field. With its 
set of four gratings and a pixel scale that can be varied by a factor
of ten, SPIFFI provides high flexibility, and at the same time 
offers the unique possibility of diffraction limited imaging spectroscopy 
at an 8m-class telescope, when fed by the adaptive optics system 
MACAO. We outline the scientific drivers for building such an instrument,
the concept of image slicing, the optical design, and the implementation
of SPIFFI.  

\end{abstract}


\section{INTRODUCTION}

\subsection{The need for near infrared integral field spectroscopy}

A large number of astronomical problems demand spectroscopy of every
image point of a two-dimensional field on the sky, because the objects
of interest have a complex spatial structure. These objects span the
whole Universe ranging from high redshift galaxies to the moons and
planets in our solar system. Merging galaxies in the early Universe
need to be spatially resolved to uncover the underlying dynamics. The
analysis of giant black holes in the center of active galactic nuclei
requires the knowledge of the two dimensional velocity fields to
overcome ambiguities from anisotropic motion. The starforming history
of extragalactic starbursts can only be derived, if we can resolve the
individual clusters. New stars are born in a highly complex
environment, exhibiting discs and outflows with different spectral
characteristics, which would be veiled in long slit spectra. There are
several classic techniques like slit-scanning,
Fabry-Perot-interferometry and Fourier-transform-spectroscopy which
provide spectra for a two dimensional field, but they all need several
exposures for a single field. However, modern integral field
spectrometers with lenslet arrays or image slicers can record the
spectra for every point of a two dimensional field in a single
exposure. Making most efficient use of the telescope, observations
with integral field spectrometers are also more easily to correct for
variations in the atmospheric transmission. This is specifically
important in the near infrared wavelength range, where the atmosphere
varies on time scales of minutes. Also adaptive optics observations,
which provide images with a spatial resolution close to the
diffraction limit of a telescope, suffer much from the time varying
correction, and advanced image reconstruction techniques like
deconvolution require a simultaneous observation of the
two-dimensional field with a field spectrometer.

There are several reasons, both object inherent and technical, to
carry out astronomical observations at near infrared wavelengths:
First, many of the faint objects we are looking for --- like in the
Hubble Deep Field --- are at high redshift.  Therefore a lot of the
well established ``optical'' spectral diagnostics are shifted beyond 1
micron.  Second, many of the interesting objects in the universe ---
like nuclei of galaxies, star and planet forming regions --- are
hidden behind dust. For example our Galactic Center is dimmed in the
visible by about 30 magnitudes, while we suffer from only 3 magnitudes
of extinction in K-Band (2.2 $\mu$m).  And third, high angular
resolution through the earth's atmosphere is much easier achieved at
longer wavelengths.  Even though there is no principle limitation to
reach the diffraction limit in the visible, the high complexity of an
adaptive optics system for very large telescopes suggests that we
start with the easier task of correcting in the near infrared.

\subsection{The SPIFFI project} 

Based on the scientific and technical motivation outlined in the
previous section, the Max-Planck-Institut f\"ur extraterrestrische
Physik (MPE) in Garching, Germany, started in the late 90's the
development of SPIFFI ({\bf SP}ectrometer for {\bf I}nfrared {\bf
F}aint {\bf F}ield {\bf I}maging), a state of the art adaptive optics
assisted near infrared integral field spectrometer. Because of the
great success of SPIFFI's precursor 3D \cite{weitzel96}, SPIFFI was
thought as a travelling instrument for several telescopes, including
the Calar Alto Observatory, the European Southern Observatory (ESO)
and the Large Binocular Telescope (LBT). In order to keep SPIFFI's
dimensions and weight to its absolute minimum, we started the
development of a new image slicer based on flared fibers
\cite{tecza98}. However, given the very good collaboration with ESO on
previous instrument projects (SHARP I and SHARP II) \cite{hofmann95},
MPE decided to build a VLT specific instrument SPIFFI, to be assisted
with ESO's adaptive optics system MACAO. This combination of an
adaptive optics system and an integral field spectrometer is jointly
refered to as SINFONI \cite{thatte98} ({\bf SIN}gle {\bf F}aint {\bf
O}bject {\bf N}ear {\bf I}nfrared {\bf I}nvestigation).  From the very
beginning, SINFONI was thought to be a fast track instrument.  First
light is foreseen in 2002 on VLT UT3. Since SPIFFI will remain at the
VLT, the constraints on size and weight have been relaxed, and a
mirror based image slicer could be considered for SPIFFI again.  Given
the superior performance of a mirror based image slicer, the tight
schedule for SPIFFI, and unexpected delays in the development of the
fiber based image slicer, we decided to equip SPIFFI with a mirror
slicer \cite{tecza00}.
\begin{figure}
\begin{center}
\begin{tabular}{c}
\psfig{figure=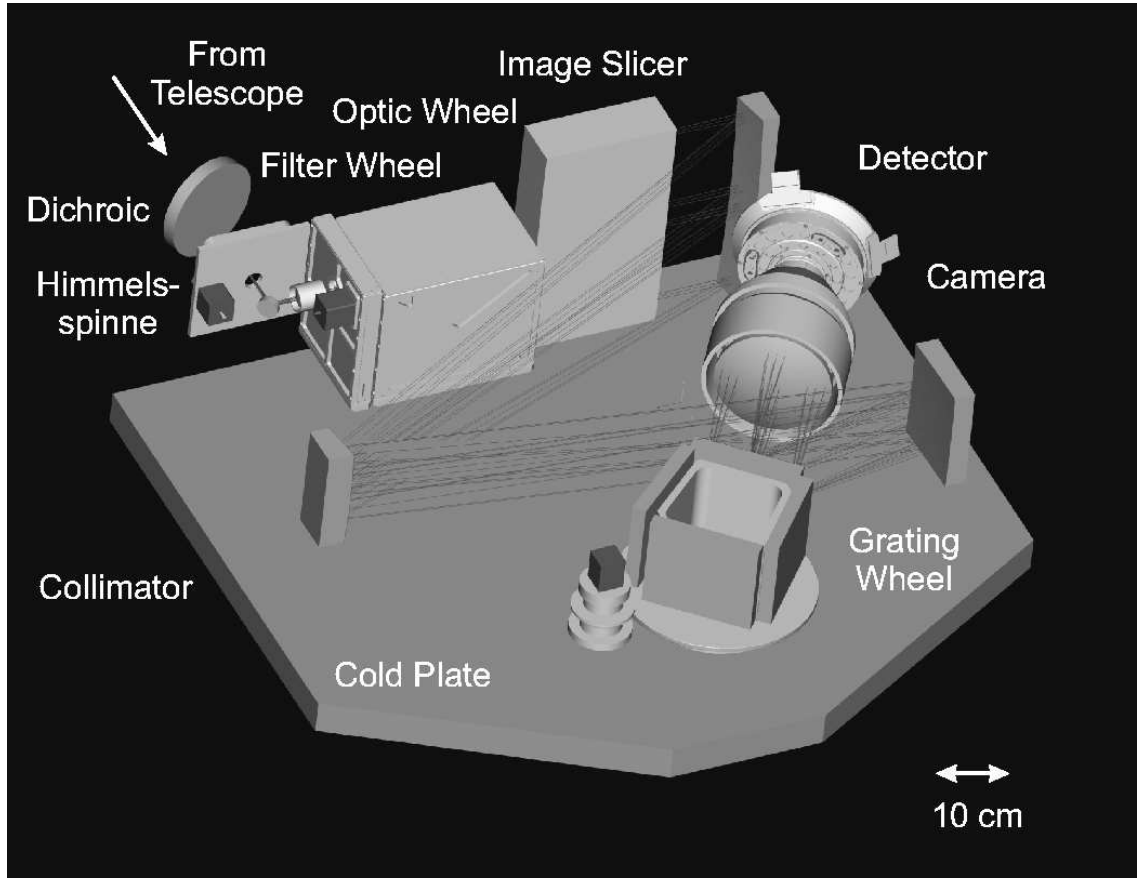,width=15cm}
\end{tabular}
\end{center}
\caption[Assembly Drawing (Perspective view)] {
   \label{assembly_perspective_bw} Assembly drawing of SPIFFI
   (perspective view): The drawing shows the main components of
   SPIFFI. The light enters the instrument from the upper left. All
   components are cooled with a cryostat (not shown) to the
   temperature of 77 K. }
\end{figure} 

\subsection{The SPIFFI instrument}

SPIFFI records simultaneously the spectra of all 32 x 32 image points
of a two-dimensional field of view. It thus allows spectroscopy of
objects with a complex spatial structure, and will make the most
efficient use of observing time compared to alternative imaging
spectrographs. Given the relevance of near infrared spectroscopy for
many areas of modern astronomy, SPIFFI will cover the wavelengths of
the three atmospheric bands J (1.1 $\mu$m - 1.4 $\mu$m), H (1.45
$\mu$m - 1.85 $\mu$m), and K (1.95 $\mu$m - 2.45 $\mu$m), i.e. from
1.1 $\mu$m - 2.45 $\mu$m.  The instrument is fully cryogenic, and will
be equipped with an 1k x 1k HAWAII \cite{hodapp96} array from
Rockwell. The image scale of SPIFFI allows both Nyquist sampled
imaging at the diffraction limit of the telescope ( 0.025
arcsec/pixel), and seeing limited observations (0.25 arcsec/pixel). An
intermediate image scale provides a compromise of field size and
spatial resolution.  The spectral resolution of the spectrometer is
about 4000 for all three wavelength bands J, H, and K, which allows
detailed kinematic study of galaxies, and at the same time an
effective OH-avoidance of the atmospheric emission lines in the NIR.
A more moderate resolution R $\approx$ 2000 is implemented, too, for
objects that are too faint for high resolution spectroscopy, covering
H \& K simultaneously. The optics is designed for gratings with a
resolution of up to 10000, which may be integrated in future upgrades
of SPIFFI.  Figure \ref{assembly_perspective_bw} shows a perspective
view of the main components of SPIFFI.

\begin{figure}
\begin{center}
\begin{tabular}{c}
\psfig{figure=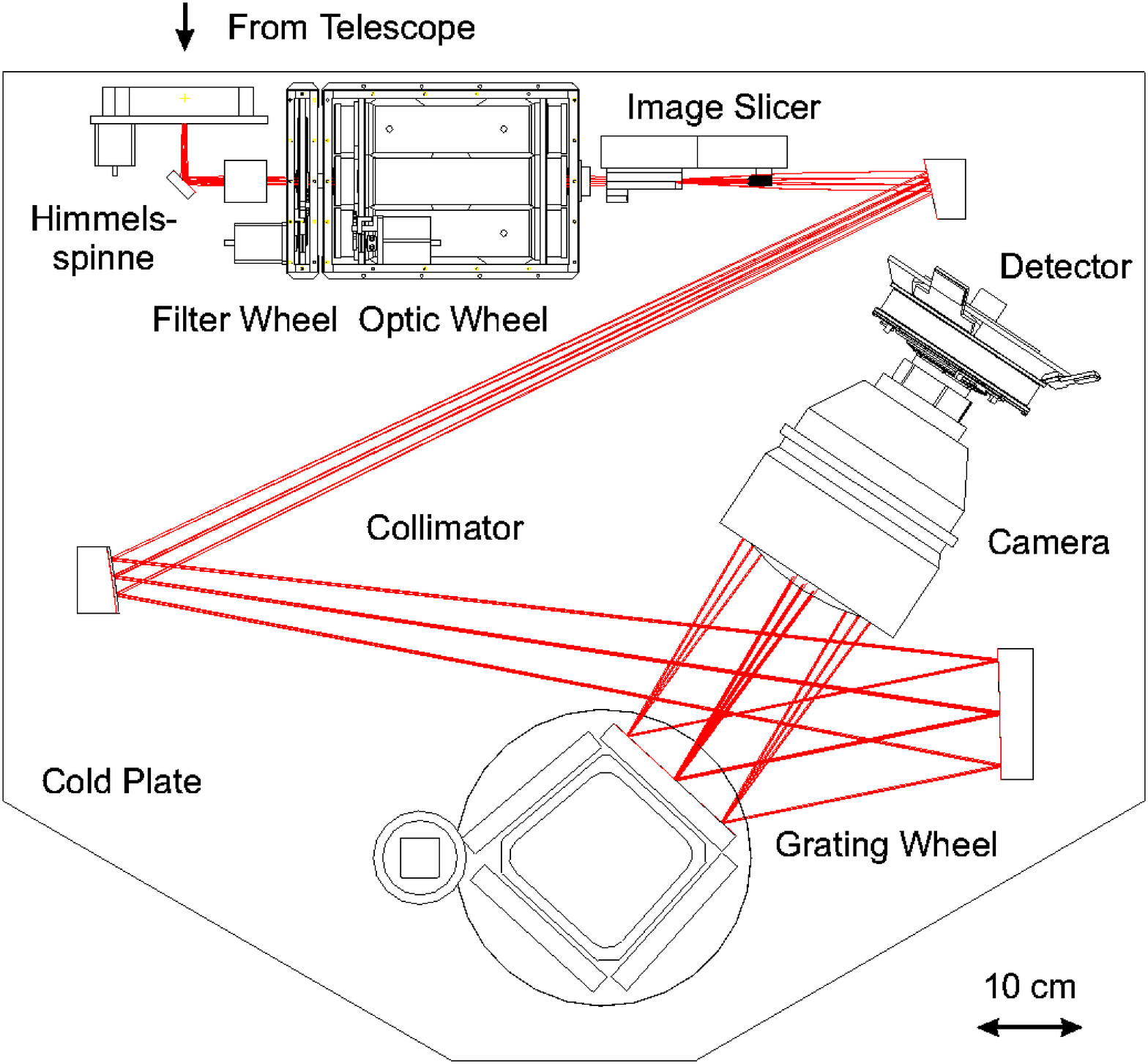,width=14cm} 
\end{tabular}
\end{center}
\caption[Assembly Drawing (Top view)] 
   { \label{assembly_top}	  
   Schematic drawing of SPIFFI (top view): The drawing shows the main
   components of SPIFFI. The light enters the instrument from the
   top.} 
\end{figure} 

\section{OPTICS}

\subsection{Overview}

The SPIFFI integral field spectrometer will consist of three basic
components: (a) Preoptics: The Preoptics reimage the object plane
from the adaptive optics onto the image slicer, providing the
different pixel scales. In addition, a cold stop at the intermediate
pupil position will allow efficient suppression of the thermal
background. The pre optics will also host broad band filters for
selecting the wavelength range. (b) Image slicer: The image slicer
cuts the two dimensional field into a set of 32 individual slitlets,
and rearranges them to a one-dimensional pseudo long slit. This image
slicing is done by a set of 64 plane mirrors. (c) Spectrometer: The
spectrometer reimages the pseudo long slit from the image slicer onto
the detector. The gratings are located at the intermediate pupil
position. In addition, a so called "Himmelsspinne" (sky spider) at the
entrance of the instrument will allow the observation of the sky
background simultaneous with the astronomical target. Figure
\ref{assembly_top} shows a schematic view of the SPIFFI optics.

\subsection{Imaging optics}

\begin{figure}
\begin{center}
\begin{tabular}{c}
\psfig{figure=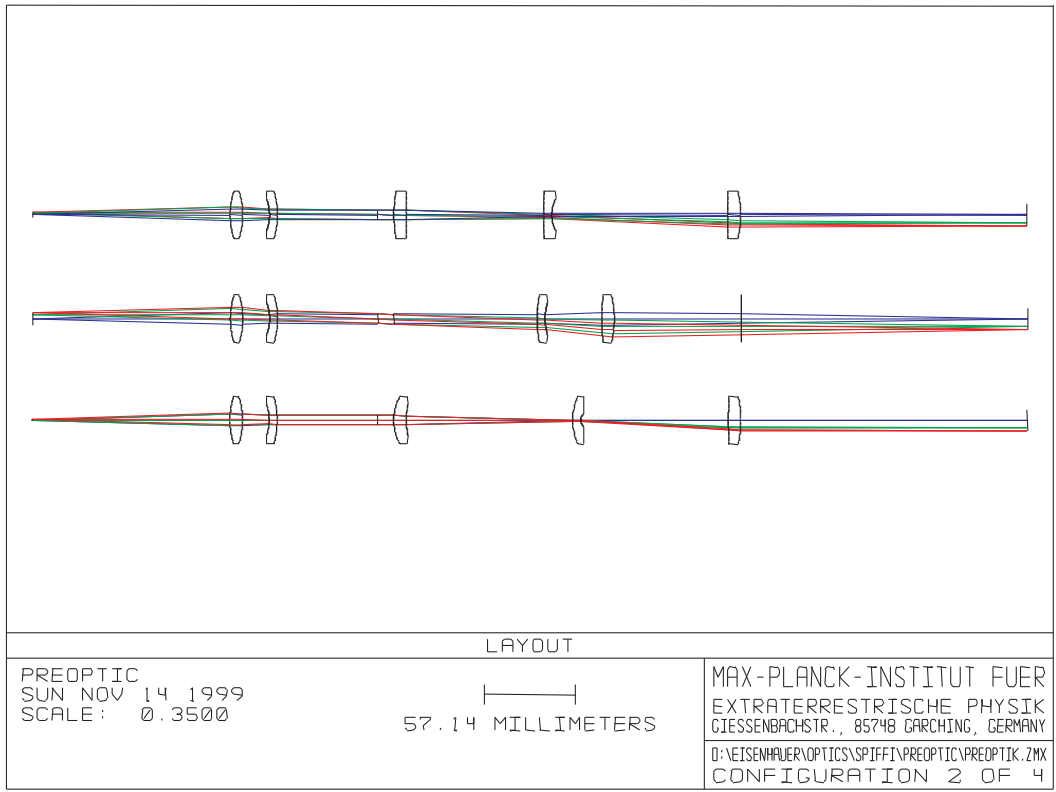,width=14cm}
\end{tabular}
\end{center}
\caption[Preoptic] { \label{preoptic} Preoptic of SPIFFI: The drawing
   shows the preoptics for the 0.1 arcsec/pixel scale (top), 0.25
   arcsec/pixel scale (middle) and 0.025 arcsec/pixel scale
   (bottom). The pupil imaging lens is omitted. }
\end{figure}

The preoptics reimage the focal plane of the adaptive optics onto the
image slicer. It consists of a fixed collimator, a filter wheel, and
an optic wheel. The preoptics provide three different magnifications
(17.8 x, 4.45 x, 1.78 x) with equivalent pixel scales of 0.025 arcsec/pixel,
0.1 arcsec/pixel, and 0.25 arcsec/pixel. In addition, a pupil imaging lens
will allow for an accurate alignment of SPIFFI's optical axis with the
telescope / adaptive optics. Figure \ref{preoptic} shows the
preoptics with the three imaging lenses.

The preoptics use only spherical lenses made from barium fluoride
and IRG2 from Schott. With a focal length of 85.5 mm, the collimator
reimages the entrance pupil on a 6.0 mm diameter cold stop. The
preoptics are telecentric. The maximum deviation from the on-axis
pupil location is less than 1 \% of the corresponding radius across
the whole field.  The residual distortion is less than 0.025 \% for
all image scales. All image scales are diffraction limited, and
provide a nominal Strehl ratio $>$ 97 \% for all wavelengths across the
whole field. The filter wheel is located in the collimated beam in
front of the cold stop.

\subsection{Image Slicer}

The image slicer \cite{tecza00} transforms the two-dimensional field
into a one dimensional slit. This long slit is then dispersed by a
classical spectrometer. The first concept of SPIFFI was based on
optical fibers \cite{tecza98}, where the image plane is sampled by a
bundle of these fibers, which are then rearranged to a
"long slit". While this concept is still promising for future projects
\cite{eisenhauer00}, the tight schedule for SPIFFI and problems with
the manufacturing of this fiber bundle led to the revival of the
mirror slicing concept, as used in the MPE 3D spectrometer
\cite{weitzel96}.  The first set of mirrors, called "Small Slicer", is
located in the object plane, and redirects the light from different
field points towards different directions, thus separating the beams
from each other. At the location of the second set of mirrors, called
"Large Slicer", the beams are completely separated. However, the beams
point towards different directions, or in other words, the exit pupils
of the different rows do not overlap anymore. The large slicer
realigns the pupils again, e.g. the beams from the different rows are
parallel at the exit of the image slicer, but still spatially
separated. Only plane mirrors are used in the image slicer.  There are
several points to consider for the detailed design of this slicer:

(a) For a total of 32 x 32 pixels, the length of the final pseudo long
slit is 1024 times the width of the small slicer mirrors. For example,
when using 1 mm wide mirrors, the spectrometer would have to work with
a slit 1 m long. In order to keep the instrument size small, one
therefore should make the mirrors as small as possible.  Choosing too
small a mirror size, however, leads to strong defocus effects,
limiting the efficiency of the small slicer. The reason for this is
that for a given projected pixel size on the sky, smaller mirrors
imply a faster beam at the slicer. A faster beam, however, reduces the
focal depth, and vignetting at the interface between two mirrors gets
more serious. Also it is very hard to manufacture mirrors smaller than
a few hundred microns.  The compromise for SPIFFI is a width of 300
$\mu$m for the mirrors of the small slicer, thus providing a pseudo
long slit of 307.2 mm.

(b) Since the beams are diverging towards the large slicer, the
light from neighboring rows will overlap slightly at the edge of the
mirrors from the large slicer, if they are aligned along a single
layer. To prevent this, the mirrors of the large slicer are aligned in
a brick wall pattern. There is no crosstalk at the position of the
large slicer, but the pseudo long slit will consist of two truncated
slits.

\begin{figure}
\begin{center}
\begin{tabular}{c}
\psfig{figure=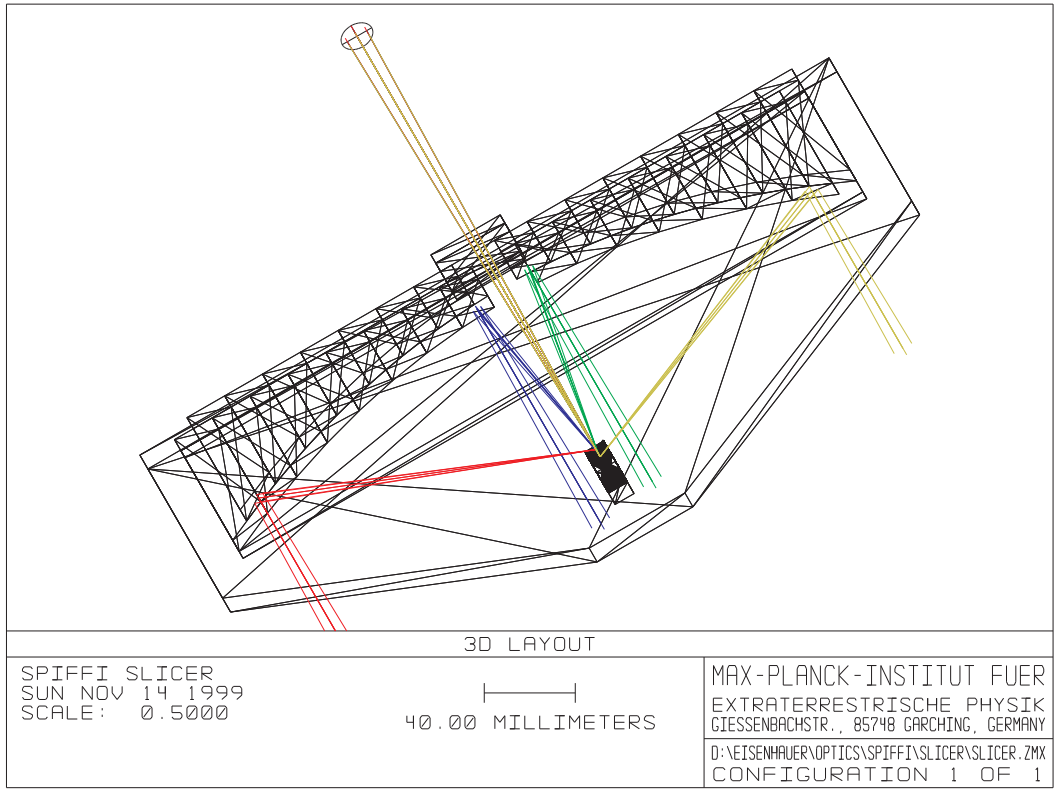,width=14cm}
\end{tabular}
\end{center}
\caption[Image Slicer] { \label{slicer} 
   SPIFFI Image slicer} 
\end{figure}

Figure \ref{slicer} shows the design for the SPIFFI mirror slicer. The
whole slicer will be fabricated from Zerodur. All elements will be
optically contacted, no glue etc. is used. The monolithic design
allows the slicer to be cooled down to the temperature of 77
K. Several cool downs have been carried out with a test slicer without
any problems. The mirrors will be coated with a highly reflective 
gold layer. 

\subsection{Spectrometer}

The spectrometer consists of a collimator, the gratings, and a
camera. 

The collimator of the SPIFFI spectrometer is a
three mirror anastigmat. It consists of one spherical and two off-axis
prolate elliptical mirrors (see figure \ref{assembly_top}).
This system combines the advantage of a large one-dimensional field of
$6.092^\circ$ (slit length: 307.2 mm) with a very compact design, and
requires only 3 mirrors. Its nominal focal length is 2886.5 mm. The
optics is operated $11.5^\circ$ off-axis. The nominal Strehl ratio is
$>$ 96 \% across the whole field, for all wavelengths, and for all pixel
scales. The location of the exit pupil for the different field points
of the long slit varies by less than 1 \% of its radius. The
distortion along the slit less than 0.2 \%.  The mirrors will be
diamond turned from nickel plated aluminum. Postpolishing will reduce
the residual surface roughness to less than 50 \AA ngstr\o m. The mirrors
will be gold coated (with protection layer).

\begin{figure}
\begin{center}
\begin{tabular}{c}
\psfig{figure=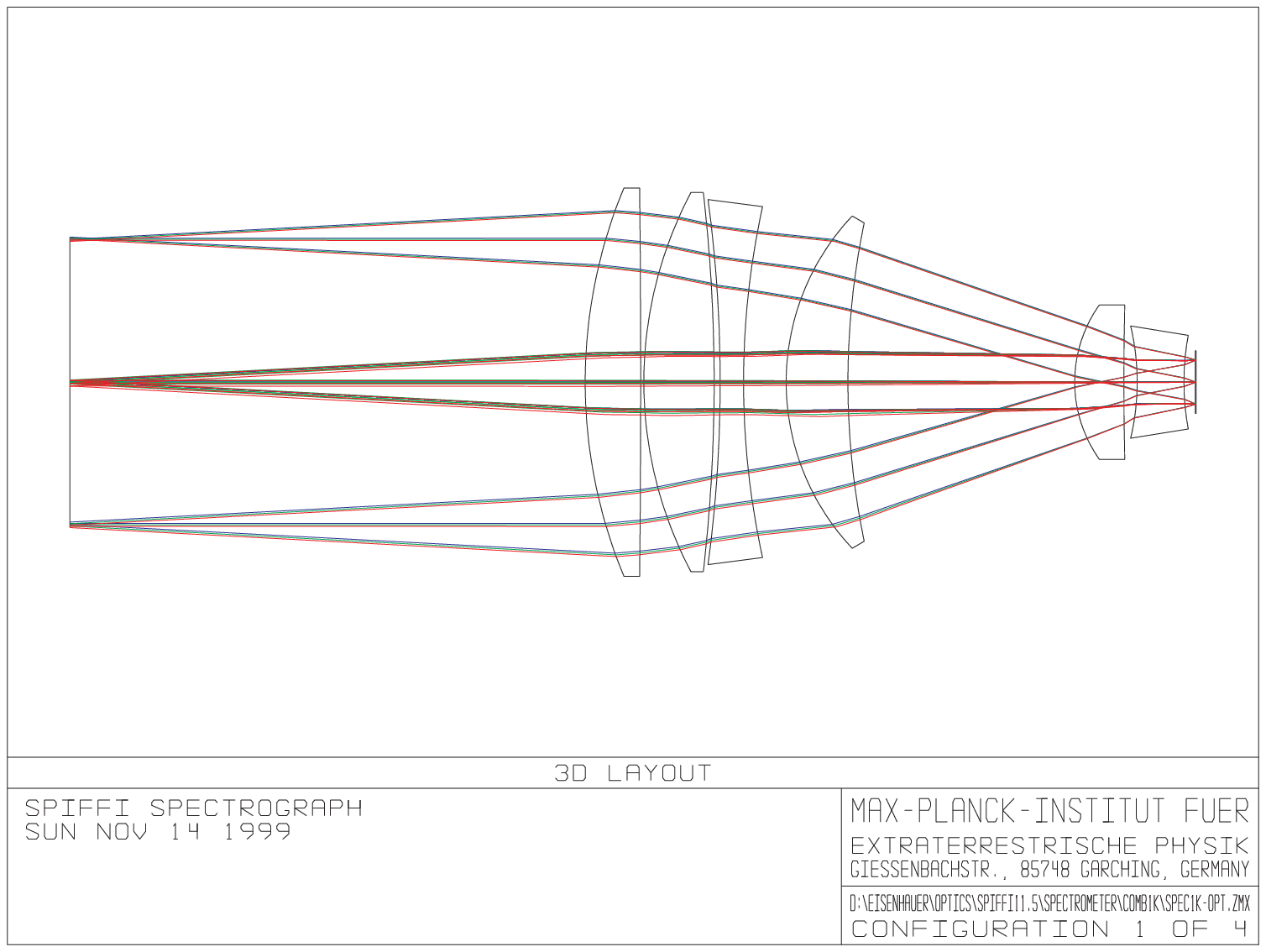,width=14cm}
\end{tabular}
\end{center}
\caption[Camera] { \label{camera} 
   Spectrometer camera of SPIFFI}
\end{figure} 

The nominal focal length of the camera (figure \ref{camera}) is 178
mm. Because of the anamorphic magnification from the grating, and
because of a small pupil distortion from the collimator, the beam size
at the exit of the grating is $\approx$ 125 mm at the 0.25
arcsec/pixel scale. The field of view is $8.615^\circ$. In order to
clear the beam, the entrance pupil of the camera was chosen to be 220
mm in front of the first lens surface. The camera was optimized to
correct for -0.2652 \% of anamorphic distortion from the gratings, so
that the spectra of a single image point differ by less than 1/5 of a
pixel from a straight line. The camera is completely diffraction
limited --- i.e. nominal Strehl ratio 100 \% --- at all wavelengths
(1.1 $\mu$m - 2.45 $\mu$m) when operated with the adaptive optics
image scale of 0.025 arcsec/pixel. The nominal Strehl ratio for the
0.1 arcsec/pixel scale is everywhere $>$ 90 \%. For operation with the
seeing pixel scale with 0.25 arcsec/pixel (biggest beam), the camera
was optimized for smallest spot size The nominal RMS spot radius is
$<$ 5 $\mu$m everywhere. For comparison, the size of a pixel is
18.5 $\mu$m.

The heavy constraints on the image quality, and the large f-number of
$\approx$ 1.4 of the camera can only be fulfilled with the use of
lenses made from the special infrared glass IRG2 from Schott. Since
this glass is not available from stock, MPE has ordered a custom melt.

SPIFFI will have four plane gratings, covering the atmospheric J, H,
K, and H \& K bands. The gratings are operated in Ebert configuration.
The size of all gratings is 160 mm x 140 mm. The blanks are made from
nickel- and gold plated aluminum and are light weighted.  All gratings
are directly ruled and are blazed for highest efficiency. The tolerance
on surface flatness corresponds to 90 nm RMS in the diffractive
wavefront at normal incidence. The average efficiency of all gratings 
will be around 60 \% to 80 \%.

\subsection{Himmelsspinne}

The Himmelsspinne facilitates simultaneous observations of
an object and a blank sky field. The observer may choose one of four
fields, with separations of about 0.25, 0.5, 0.75 and 1 arc minute
from the center of the SPIFFI field of view. Light from the chosen
field is redirected via two plane mirrors onto a corner of the image
slicer, providing simultaneous measurements of the night sky
background level. The number of pixels occupied by the sky fields can
be selected by the observer.

\pagebreak

\section{MECHANICS}

\subsection{Cryostat} 

All optics and the detector of SPIFFI will be operated at 77 K in a
liquid nitrogen bath cryostat. The cold volume is approximately 1100
mm x 950 mm x 450 mm. The outer dimensions of the cryostat are
approximately 1300 mm x 1150 mm x 1050 mm. The reservoir holds a
maximum of 47 liters of liquid nitrogen, and is oversized by a factor
of two, so that the cryostat can be tilted by 90$^\circ$ for operation
at the Cassegrain focus of the VLT.  We estimate the total heat
dissipation to be about 42 W. The hold time of the cryostat will be at
least 36 hours.

\subsection{Moving mechanisms}

The cryostat will incorporate four different moving mechanisms, to
serve the following functions: 

(a) A grating wheel allows a choice of one of the four installed
gratings. (b) A filter wheel to choose the appropriate band pass
corresponding to the selected grating. (c) A scale changer with four
different positions, so as to choose one of three pixel scales or the
pupil imaging optics for pupil alignment. (d) A sky field selector for
the Himmelspinne, allowing the observer to choose one of four
predefined sky field points for simultaneous observation of the sky
background.

All cryogenic moving assemblies will be outfitted with 
commercial stepper motors, which are modified for cryogenic operation.

\section{ELECTRONICS}

\subsection{Detector readout electronics}

SPIFFI is equipped with a 1k x 1k HAWAII array from Rockwell. The
hybrid structure of the arrays allows for a non destructive readout of
the individual pixels, so that advanced multiple read modes can be
applied for efficient noise reduction. The electronics used for SPIFFI
will be the ESO IRACE \cite{meyer98} system. 

\subsection{Instrument control electronics}

\begin{figure}
\begin{center}
\begin{tabular}{c}
\psfig{figure=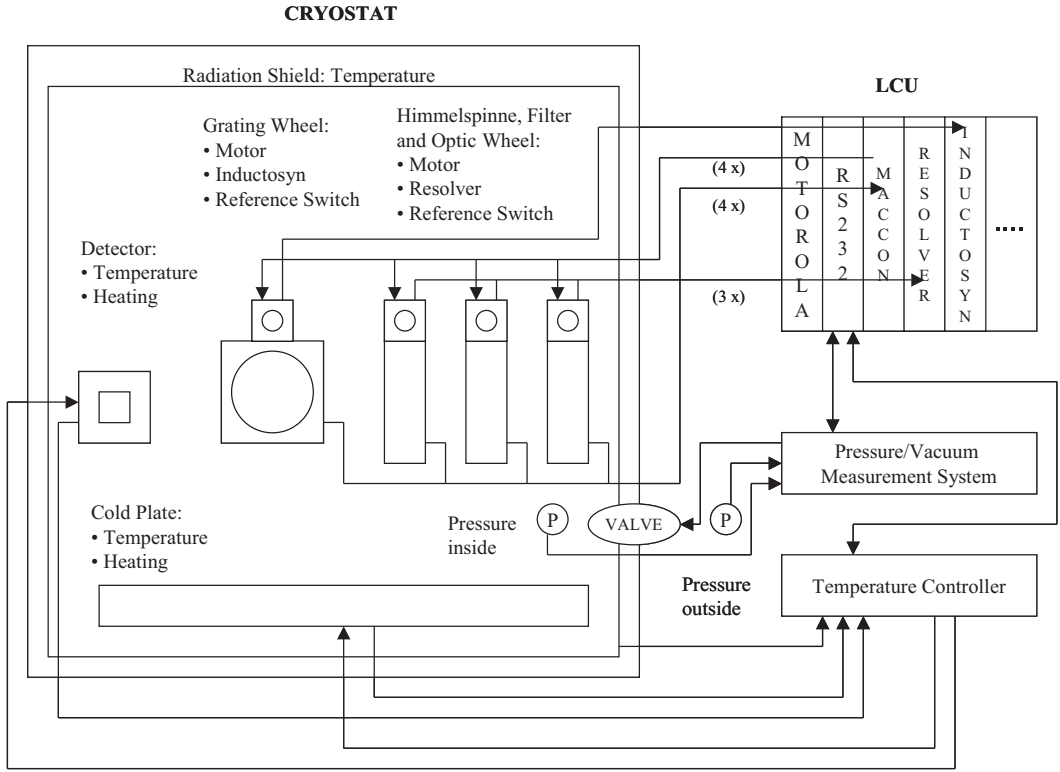,width=14cm}
\end{tabular}
\end{center}
\caption[Instrument electronic] { \label{electronic}	 
   Overview of the instrument control electronics}
\end{figure}

The instrument control electronics governs the function of all moving
mechanisms within the cryostat.  An overview of the instrument control
electronics is shown in figure \ref{electronic}. A motor controller
controls the four cryogenic motors which allow the observer to choose
the grating, choose the filter, select an appropriate spatial scale,
and select a sky setting for the Himmelspinne. The grating wheel is
equipped with a high resolution cryogenic encoder with 2 arc second
angular resolution.  The encoder signal is fed back to the motor
controller, which can position the gratings with an accuracy better
than 0.2 pixels. The operator can offset the spectra on the detector
by 1/2 pixel, for proper Nyquist sampling of the spectra.  The motors
from the Himmelsspinne, the optic wheel and the filter wheel are
equipped with resolvers. All cryogenic moving mechanisms are fitted
with reference switches whose signals are fed back to the motor
controller. In addition, the instrument control electronics includes
temperature sensors which measure the temperatures inside the cryostat
at various critical locations. The readout from the vacuum gauges is
also reported to the control computer.

\section{Instrument Performance}

The SPIFFI instrument is designed to maintain a very high optical
throughput ($\approx$ 32 \%) at wavelengths of 1.1 $\mu$m to 2.45
$\mu$m. Since all optics, including the image slicer, is cooled down
to 77 K, the thermal background from the instrument is negligible
compared to the contribution from the telescope and the sky. While the
K-band instrument performance is mostly limited by the thermal
emission from the telescope and the sky, J- and H-band observations
are limited by the specifications of the Rockwell HAWAII detector. The
final sensitivity of SPIFFI will highly depend on the seeing
conditions, the performance of the adaptive optics system MACAO, and
the brightness of the sky, especially the OH lines, during
observation. Therefore the following numbers (tables \ref{seeing} and
\ref{ao}) should only give the reader a rough indication of the
sensitivity of SPIFFI when operated under normal seeing conditions,
and when taking advantage of the adaptive optics.

\begin{table}   
\caption{Sensitivity of SPIFFI for seeing limited observations} 
\label{seeing}
\begin{center}       
\begin{tabular}{|l|l|l|l|l|}
\hline
 & J-band & H-band & K-Band & H\&K-band \\
\hline
Point Source Sensitivity [mag] & 21.4 & 20.7 & 18.3 & 19.3 \\
Surface Brightness Sensitivity [mag/arcsec$^2$] & 20.2 & 19.5 & 17.1 &
 18.1 \\

\hline
\end{tabular}
\end{center}
\end{table}

\begin{table}
\caption{Sensitivity of SPIFFI for adaptive optics assisted observations} 
\label{ao}
\begin{center}       
\begin{tabular}{|l|l|l|l|}
\hline
 & H-band & K-Band & H\&K-band \\
\hline
Point Source Sensitivity [mag] & 20.7 & 19.6 & 20.5 \\
Surface Brightness Sensitivity [mag/arcsec$^2$] & 15.0 & 14.2 &
 15.0 \\
\hline
\end{tabular}
\end{center}
\end{table}

The sensitivity numbers were calculated for a spectral resolution of R
= 4200 in J- and H-band, R = 4400 in K-band and R = 1900 in H\&K-band,
on source integration time of 1 hour, signal-to-noise ratio of 3 per
spectral channel, a dark current of 0.1 electrons/s and a
read noise of 8 electrons. The numbers for H-band observations include
OH-avoidance techniques. For typical seeing limited observations we
assume the large image scale with 0.25 arcsec / pixel image scale and
a seeing where 50 \% of a point source flux is within 0.7 arcsec.
The sensitivity numbers for adaptive optics assisted observations
were derived for the small image scale with 0.025 arcsec / pixel 
and a Strehl ratio of 50 \%.


\begin{thebibliography}{99}

\bibitem{weitzel96} Weitzel L., Krabbe A., Kroker H., Thatte N.,
Tacconi-Garman L. E., Cameron M., and Genzel R., ``3D: The next
generation near-infrared imaging spectrometer'', Astronomy and
Astrophysics Supplement Series, 119, 531, 1996
\bibitem{tecza98} Tecza M., Thatte N. A., Krabbe A., and
Tacconi-Garman L. E., "SPIFFI: a high-resolution near-infrared
imaging spectrometer", Proceedings of SPIE, 3354, 394, 1998
\bibitem{hofmann95} Hofmann R., Brandl B., Eckart A., Eisenhauer
F., and Tacconi-Garman L. E., ``High-angular-resolution NIR astronomy
with large arrays (SHARP I and SHARP II)'', Proceedings of SPIE, 2475,
192, 1995
\bibitem{thatte98} Thatte N., Tecza M., Eisenhauer F. et al.,
``SINFONI: a near-infrared AO-assisted integral field spectrometer for
the VLT'', Proceedings of SPIE, 3353, 704, 1998
\bibitem{tecza00} Tecza M., Thatte N., Eisenhauer F., Mengel S.,
R\"ohrle C., and Bickert K., ``The SPIFFI image slicer: Revival of
image slicing with plane mirrors'', Proceedings of SPIE, accepted,
2000
\bibitem{hodapp96} Hodapp K.-W., Hora J. L., Hall D. N. et al.,
``The HAWAII Infrared Detector Arrays: testing and astronomical
characterization of prototype and science-grade devices'', New
Astronomy, 1, 177, 1996
\bibitem{eisenhauer00} Eisenhauer F., Tecza M., Thatte N., Mengel
S., Hofmann R., and Genzel R., ``Near-Infrared-Spectroscopy with
Extremely Large Telescopes: Integral-Field- versus
Multi-Object-Instruments'', ESO Conference and Workshop Proceedings,
57, 292, 2000
\bibitem{meyer98} Meyer M., Finger G., Mehrgan H., Nicolini G., and
Stegmeier J., ``ESO infrared detector high-speed array control and
processing electronic IRACE'', Proceedings of SPIE, 3354, 134, 1998

\end{thebibliography}
\end{document}